\documentclass{iau}
\usepackage{amsmath}

\title[The Pulsation-Rotation Interaction] 
{The Pulsation-Rotation Interaction: Greatest Hits and the B-Side}

\author[Townsend]   
{Rich Townsend}

\affiliation{%
Department of Astronomy, University of Wisconsin-Madison, Madison, WI 53706, USA}

\pubyear{2013}
\volume{301}  
\pagerange{1--7}
\jname{Precision Asteroseismology}
\editors{W. Chaplin, J. Guzik, G. Handler \& A. Pigulski}
\begin{document}

\maketitle

\begin{abstract}
It has long been known that rotation can have an appreciable impact on
stellar pulsation --- by modifying the usual p and g modes found in
the non-rotating case, and by introducing new classes of
mode. However, it’s only relatively recently that advances in
numerical simulations and complementary theoretical treatments have
enabled us to model these phenomena in any great detail. In this talk
I'll review highlights in this area (the `Greatest Hits'), before
considering the flip side (or the `B-side', for those of us old enough
to remember vinyl records) of the pulsation-rotation interaction: how
pulsation can itself influence internal rotation profiles.
\end{abstract}

\firstsection 

\section{Introduction}

Back at the dawn of civilization, the principal medium for
distributing music was the vinyl gramophone record. Those who grew up
in the vinyl era will recall that a record had two sides --- the `A'
side featuring the hit(s) that usually motivated the initial purchase
of the record, and the `B' side which contained somewhat more-esoteric
material often destined to languish in obscurity\footnote{There are,
  of course, exceptions: \emph{Rock Around the Clock} was first
  released by Bill Haley \& His Comets (a very astronomical band!) on
  the B-side.}

I bring up these facts to draw a strained analogy to the
pulsation-rotation interaction in stellar astrophysics. The `A' side
with which we're all familiar comprises the effects of rotation on
pulsation; but there's also an accompanying `B' side which considers
the influence that pulsation might have on rotation, and indeed the
host star's overall evolution. In this contribution I first review the
`greatest hits' on the `A' side
(\S\S\ref{sec:perturb}--\ref{sec:differential}), before highlighting
some important developments from the `B' side (\S\ref{sec:transport}).

\section{Perturbative Approaches} \label{sec:perturb}

\cite[Ledoux (1949)]{Led1949} and \cite[Cowling \& Newcomb
  (1949)]{CowNew1949} first considered the effects of slow rotation on
the oscillation frequencies of a star. As seen from an inertial frame,
these can be expressed as
\begin{equation} \label{eqn:linear-split}
\omega = \omega_{0} + m \Omega (1 - C_{n,\ell}),
\end{equation}
where $\omega_{0}$ is the frequency the mode would have in the absence
of rotation, $\Omega$ is the rotation angular frequency (for now,
assumed uniform), $C_{n,\ell}$ the Ledoux constant and $n,\ell,m$ are
the usual mode radial order, harmonic degree and azimuthal order,
respectively. The Ledoux constant accounts for the effects of the
Coriolis force on the mode. It is usually positive, because the
Coriolis force tends to counteract the restoring force on displaced
fluid elements for prograde modes ($m > 0$), leading to smaller
frequencies (vice-versa for retrograde modes with $m <
0$). (Sometimes, however, phenomena such as mode coupling can produce
negative $C_{n,\ell}$; Takata \& Saio, these proceedings, present an
example of this). The other term in the parentheses accounts for the
Doppler shift in transforming from the co-rotating frame to the
inertial frame.

The frequency splitting described by eqn.~(\ref{eqn:linear-split}) is
linear in $m$. This is entirely analogous to the Zeeman splitting of
atomic energy levels in a weak magnetic field, and the same
first-order perturbation expansion approach underpins the analysis of
both phenomena. Moving to more-rapid rotation requires a higher-order
perturbation expansion. \cite[Simon (1969)]{Sim1969} and a number of
subsequent authors extended the formalism to second order in $\Omega$,
and \cite[Soufi et al. (1998)]{Sou1989} took it to third order;
however, these treatments are significantly more complicated than
eqn.~(\ref{eqn:linear-split}). The value of going to even high orders
is moot, because the complexity of the problem becomes unmanageable;
equally importantly, there's a point where the effects of rotation can
no longer be considered a small perturbation to the non-rotating
pulsation equations. Then, so-called `non-perturbative' approaches are
required.

\section{Non-Perturbative Approaches} \label{sec:non-perturb}

The pulsation equations in a rotating star comprise a 2-dimensional
boundary value problem (BVP), with radius $r$ and co-latitude $\theta$
the independent variables and the frequency $\omega$ serving as an
eigenvalue. Non-perturbative approaches to solving these equations
fall into four main groups.

\subsection{Direct Methods} \label{ssec:direct}

Conceptually, the simplest non-perturbative approach is to approximate
the pulsation equations using finite differences on a 2-D ($r,\theta$)
grid. This leads to a large set of algebraic equations, which can be
solved using sparse-matrix algorithms. \cite[Clement (1998, plus a
  number of earlier papers)]{Cle1998} and \cite[Savonije et
  al. (1995)]{Sav1995} use direct methods, but they haven't been
more-widely adopted for reasons which aren't immediately obvious. (My
personal perspective is that the numerical aspects of direct methods
are rather daunting).

\subsection{Spectral Methods} \label{ssec:spectral}

Spectral methods expand the angular dependence of solutions as
(typically large, but finite) sums of spherical harmonics
$Y^{m}_{\ell}(\theta,\phi)$, with the same azimuthal orders but
harmonic degrees $\ell = |m|,|m|+2,|m|+4,\ldots$ for even-parity modes,
and $\ell = |m|+1,|m|+3,|m|+5,\ldots$ for odd-parity modes. This
reduces the pulsation equations to a 1-D BVP, with the expansion
coefficients being the unknowns. This BVP is solved using the same
general techniques as in the non-rotating case (see, e.g., Townsend \&
Teitler 2013, and references therein), although the computational cost
is much higher.

Spectral methods are the oldest of the approaches described here,
dating back to the pioneering work on the oscillations of rotating
polytropes by \cite[Durney \& Skumanich (1968)]{DurSku1968}. They have
become increasingly popular in recent years (e.g., \cite[Lee \&
  Baraffe 1995]{LeeBar1995}; \cite[Reese et al. 2006]{Ree2006};
\cite[Ouazzani et al. 2012]{Oua2012}), perhaps driven by the advent of
inexpensive high-performance computing hardware. One criticism leveled
at spectral methods is that the truncation of the spherical-harmonic
expansion necessarily makes them approximate. This is technically
true, but only inasmuch as \emph{any} numerical solution of a system
of differential equations is an approximation. The number of spherical
harmonics can always be made sufficiently large to achieve the desired
level of accuracy --- much as the grid spacing in a finite-difference
method can always be taken sufficiently small.

\subsection{Ray-Tracing Methods} \label{ssec:ray}

Ray-tracing methods treat the pulsation equations in an asymptotic
limit analogous to the geometric limit of optics. The resulting
eikonal equation is integrated using the method of characteristics (at
a computational cost much smaller than the direct or spectral methods
described above), to find the ray paths followed by short-wavelength
acoustic waves through a rotating model star. The properties of these
rays can be studied using Poincar\'e surface sections, which mark each
passage of a ray through a fixed-radius surface with a point plotted in the
$\theta$-$k_{\theta}$ (polar wavenumber) plane.

Ligni\`eres \& Georgeot (2008) use ray tracing to show that the
acoustic properties of rapidly rotating polytropes fall into three
main groups: rays bouncing internally between mid-latitude surface
regions in the northern and southern hemispheres, rays confined to a
surface layer at all latitudes, and rays completely filling the
interior. These groups correspond, respectively, to the three classes
of global p-mode explored by Reese et al. (2009) using a spectral method:
 island modes (small $\ell-|m|$), whispering gallery modes (large
$\ell-|m|$), and chaotic modes (intermediate $\ell-|m|$).

\subsection{Traditional Approximation} \label{ssec:trad}

The traditional approximation neglects the horizontal component of the
angular velocity vector when evaluating the Coriolis
force. Originating in the geophysical literature \cite[(see Eckart
  1960)]{Eck1960}, it is a reasonable approximation in radiative
regions when both the oscillation frequency and the rotation frequency
are much smaller than the local Brunt-V\"ais\"al\"a frequency $N$ ---
that is, for intermediate- and high-order g-modes.

If the centrifugal force and the gravitational potential perturbations
are also neglected, the traditional approximation brings a huge
simplification to the pulsation equations: it allows them to be
separated in $r$ and $\theta$. Solution then proceeds as in the
non-rotating case, with only two substantive changes: spherical
harmonics $Y^{m}_{\ell}(\theta,\phi)$ are replaced by so-called Hough
functions $\Theta(\theta) \exp({\rm i}m \phi)$ (\cite[Hough
  1897]{Hou1897}; see also \cite[Townsend 2003a]{Tow2003a}), which are
the eigenfunctions of Laplace's tidal equation, and $\ell(\ell+1)$
terms are replaced by the corresponding eigenvalues $\lambda$ of the
tidal equation.

These eigenvalues depend on the `spin parameter' $\nu =
2\Omega/\omega_{\rm c}$, where
\begin{equation} \label{eqn:omega-c}
\omega_{\rm c} \equiv \omega - m\Omega
\end{equation}
is the oscillation frequency in the co-rotating frame. In the limit
$\nu \rightarrow 0$, $\lambda \rightarrow \ell(\ell+1)$; but for $\nu
\gtrsim 1$, $\lambda$ departs markedly from its non-rotating value. It
is this `inertial regime' which remains inaccessible to perturbative
approaches. (As an aside: the significance of $\nu$ is that it
measures by how much the star turns during one oscillation cycle ---
and thus to what extent a given mode is `aware' its frame of reference
is rotating).

There's an important caveat to the traditional approximation: it
neglects the possibility of resonant coupling between pairs of modes
with the same $m$. \cite[Lee \& Saio (1989)]{LeeSai1989} demonstrated
that these resonances are manifested in avoided crossings between the
mode frequencies; but in the traditional approximation the avoided
crossings are transformed into ordinary crossings, because the
resonances are suppressed. This explains why the traditional
approximation is unable to reproduce the properties of the rosette
modes found in rapidly rotating polytropes by Ballot et al. (2012); as
Takata \& Saio (these proceedings) demonstrate, these modes result
from near-degeneracies in the non-rotating frequency spectrum, which
are then pushed into resonance by the Coriolis force.

\section{The Rapidly Rotating Limit} \label{sec:rapid}

The complexity of the higher-order perturbative approaches
(\S\ref{sec:perturb}) might make us concerned that understanding
oscillations in rapidly rotating stars is going to be extremely
difficult. Fortunately, however, this often turns out not to be the
case; new regularities appear in the frequency spectra, in just the
same way that atomic energy levels become regular again in the
strong-field limit (the Paschen-Back effect).

\subsection{Regularities in p-Mode Spectra} \label{ssec:reg-p}

Based on fitting to spectral-method calculations, but also guided by
insights from ray tracing, Reese et al. (2009) propose an empirical
formula for the regularities seen in the frequency spectra of island
acoustic modes:
\begin{equation} \label{eqn:p-mode-freq}
\omega_{\rm c} \approx 
\tilde{n} \tilde{\Delta}_{\tilde{n}} + 
\tilde{\ell}\tilde{\Delta}_{\tilde{\ell}} + 
m^{2} \tilde{\Delta}_{\tilde{m}} +
\tilde{\alpha}.
\end{equation}
Here, the various $\tilde{\Delta}$ terms together with
$\tilde{\alpha}$ are obtained from least-squares fitting to the
calculated frequency spectra, while $\tilde{n}$ and $\tilde{\ell}$ are
new mode indices which correspond to the number of eigenfunction nodes
along and parallel to the ray paths stretching between the two
mid-latitude surface endpoints of island modes (see Fig. 3 of Reese et
al. 2009). These indices can be related to the radial order and
harmonic degree of the modes' non-rotating counterparts via
\begin{equation} \nonumber
\tilde{n} = 2 n, \qquad
\tilde{\ell} = \frac{\ell - |m|}{2}
\end{equation}
for even-parity modes, and by
\begin{equation} \nonumber
\tilde{n} = 2 n + 1, \qquad
\tilde{\ell} = \frac{\ell - |m| - 1}{2}
\end{equation}
for odd-parity modes.

A hand-waving narrative can be used explain the form of
eqn.~(\ref{eqn:p-mode-freq}). The terms proportional to $\tilde{n}$
and $\tilde{\ell}$ appear by direct analogy to the standard asymptotic
expression for p-mode frequencies in a non-rotating star (e.g., Aerts
et al. 2010, their eqn. 3.216), which contains terms linear in $n$ and
$\ell$. The term proportional to $m^{2}$ accounts for the bulk effects
of the centrifugal force, which do not depend on the sign of
$m$. Finally, the $\tilde{\alpha}$ term accounts for the phase of
waves at the stellar surface.

\subsection{Regularities in g-Mode Spectra}

Ballot et al. (2010) explore g modes in rotating stars using a
spectral method. Although these authors' focus is primarily on the
inadequacies of perturbative approaches in the inertial regime ($\nu >
1$), their Figs. 4 and 5 illustrate quite strikingly that, as with the
p modes above, new regularities appear in the g-mode frequency
spectrum at rapid rotation rates. 

These are a consequence of mode trapping in an equatorial
waveguide. When $\nu > 1$ the Coriolis force prevents g modes from
propagating outside of the region $|\cos\theta| \leq \nu^{-1}$. In the
limit $\nu \gg 1$, the trapping can be modeled using an asymptotic
treatment of Laplace's tidal equation first developed by \cite[Matsuno
  (1966)]{Mat1966}. The eigenvalue $\lambda$ (cf. \S\ref{ssec:trad})
is found as
\begin{equation} \label{eqn:lambda}
\lambda \approx 
\begin{cases}
\nu^{2} (2 \ell_{\mu} - 1)^{2} & \ell_{\mu} \ge 1, \\
m^{2}                         & \ell_{\mu} = 0
\end{cases}
\end{equation}
\cite[(e.g., Bildsten et al. 1996)]{Bil1996}, where the mode index
$\ell_{\mu}$ counts the number of zonal nodes in the radial
displacement eigenfunction. This index is related to the harmonic
degree and azimuthal order of the modes' non-rotating counterparts via
\begin{equation} \nonumber
l_{\mu} = \ell - |m|
\end{equation}
for prograde and axisymmetric ($m = 0$) modes, and
\begin{equation} \nonumber
l_{\mu} = \ell - |m| + 2
\end{equation}
for retrograde modes.

The co-rotating frequencies of high-order g modes depend on $\lambda$
via
\begin{equation} \nonumber
\omega_{\rm c} \approx \frac{\sqrt{\lambda}}{\pi (n + \alpha)} \int \frac{N}{r} \,{\rm d} r
\end{equation}
this is just the usual asymptotic expression (e.g., eqn. 3.235 of
Aerts et al. 2010) with $\sqrt{\ell(\ell+1)}$ replaced by
$\sqrt{\lambda}$. Combining this with eqn.~(\ref{eqn:lambda}) and
solving for $\omega_{\rm c}$ yields
\begin{equation} \label{eqn:omega-eq}
\omega_{\rm c} \approx \left[ \frac{2 \Omega (2 \ell_{\mu} - 1)}{\pi (n + \alpha)} 
\int \frac{N}{r} \,{\rm d} r \right]^{1/2}
\end{equation}
for modes with $\ell_{\mu} \ge 1$, and
\begin{equation} \label{eqn:omega-eq-kelvin}
\omega_{\rm c} \approx \frac{m}{\pi (n + \alpha)}
\int \frac{N}{r} \,{\rm d} r
\end{equation}
for the $\ell_{\mu} = 0$ modes, which are sometimes labeled equatorial
Kelvin modes \cite[(Townsend 2003a)]{Tow2003a}. These Kelvin modes
have an azimuthal phase velocity $\omega_{\rm c}/m$ that doesn't
depend on $m$; they may therefore be able to explain the unusual
uniformly spaced low-frequency modes seen in some $\delta$ Scuti stars
(e.g., KIC 8054146; see Breger, these proceedings).

These expressions indicate that a mode multiplet with a given $n,\ell$
and $-\ell \leq m \leq \ell$ will reorganize itself into $\ell + 2$
distinct frequencies, corresponding to the permitted indices $0 \leq
\ell_{\mu} \leq \ell + 1$. One corresponds to the Kelvin mode $m =
\ell$. Of the remaining $\ell+1$ distinct frequencies, the lowest
corresponds to the $m = \ell - 1$ mode, the next $\ell-1$ are made
from pairings between prograde and retrograde modes with the same
$\ell_{\mu}$, and the final, highest frequency corresponds to the
$m=-1$ mode. This is exactly the pattern seen in Figs.~4 and 5 of
Ballot et al. (2010) in the rapidly rotating limit (note that these
authors' sign convention for $m$ is reversed), confirming the analysis
here.

\subsection{Inertial Modes}

The foregoing discussion focuses on pulsation modes of rapidly
rotating stars that have oscillatory counterparts in the non-rotating
limit (in other words, $\omega_{c}^{2}$ remains greater than zero and
deforms continuously as $\Omega$ is varied between the two
limits). However, the Coriolis force introduces additional classes of
`inertial' mode whose non-rotating counterparts have $\omega^{2} \le
0$. Regions with negative $N^{2}$ can be stabilized against
perturbations with wavenumber $\mathbf{k}$ if
\begin{equation} \nonumber
N^{2} k_{\perp}^{2} > - (2 \mathbf{\Omega}\cdot\mathbf{k})^{2}
\end{equation}
(here, $k_{\perp}$ is the horizontal component of
$\mathbf{k}$). Convective fluid motions are then transformed into
oscillatory motions, with the Coriolis force serving as the restoring
force. 

The Coriolis force likewise transforms the trivial toroidal modes
having $\omega^{2} = 0$ in the non-rotating limit into r (Rossby)
modes with $\omega_{c}^{2} > 0$. The restoring force for r modes does
not depend on the stellar structure, but instead comes from
conservation of total vorticity \cite[(see Saio 1982 for an
  illuminating discussion)]{Sai1982}. In the slowly rotating limit
their frequencies are given by
\begin{equation} \nonumber
\omega_{c} = -\frac{2 m \Omega}{\ell(\ell+1)}.
\end{equation}
The opposite signs of $\omega_{\rm c}$ and $m$ in this expression tell
us that r modes are always retrograde in the co-rotating frame. Toward
more-rapid rotation the frequencies depart from this formula,
acquiring a dependence on the underlying stellar structure. The
departures are especially pronounced for the $m=-\ell$ modes, which
behave more like g modes and indeed follow the equatorial waveguide
expression~(\ref{eqn:omega-eq}) with $\ell_{\mu} = 1$. These `mixed
gravity-Rossby' modes pair up with the $m = \ell - 1$ modes of
frequency multiplets, so the number of distinct frequencies in the
multiplets remains unchanged. Due to their g-mode character they can
be excited by the $\kappa$ mechanism, as demonstrated for instance in
\cite[Townsend (2005)]{Tow2005}.

\section{Mode Visibilities}

Rapid rotation tends to reduce the photometric visibility of
oscillations.  For p modes with small $\ell-|m|$ this is a consequence
of their transformation into island modes, whose surface amplitudes
are appreciable only at mid-latitudes; for g modes, this results from
confinement in the equatorial waveguide. As a result, it becomes
challenging to detect modes photometrically in rapidly rotating stars
--- and even if modes \emph{can} be seen, they are subject to strong
selection effects (which at low frequencies strongly favor equatorial
Kelvin modes; see \cite[Townsend 2003b]{Tow2003b}). For an overview of
recent developments in this area, see \cite[Daszy\'nska-Daszkiewicz et
  al. (2007)]{Das2007} and \cite[Reese et al. (2013)]{Ree2013}.

\section{Differential Rotation} \label{sec:differential}

So far I've focused on the simple case of uniform rotation. However,
there's evidence from a multitude of sources that the internal
rotation of stars is differential in $r$ and/or $\theta$ (the most
well-known example is the Sun; see \cite[Thompson et
  al. 2003]{Tho3003}). Both perturbative and non-perturbative
approaches can readily be adapted to handle differential rotation; but
here, let's focus on an even-simpler analysis.

If the Coriolis and centrifugal force are neglected, then the only
effect of rotation is the Doppler shift in transforming between
co-rotating and inertial frames. However, in a differentially rotating
star the notion of a global co-rotating frame must be replaced by a
continuous sequence of local frames which rotate with angular
frequency $\Omega(r,\theta)$. Via equation~(\ref{eqn:omega-c}), the
co-rotating frequency is thus a function of position in the star, and
in principle can vanish wherever $m \Omega = \omega$. At these
locations, known as `critical layers' (see Mathis, Alvan \& Decressin,
these proceedings), the dispersion relation for low-frequency gravity waves
\begin{equation} \nonumber
k_{r}^{2} \approx k_{\perp}^{2} \frac{N^{2}}{\omega_{\rm c}^{2}}
\end{equation}
suggests that the radial component $k_{r}$ of the wavenumber should
diverge. In reality, what will happen is that the wavelength near a
critical layer becomes so short that significant radiative dissipation
occurs. Thus, critical layers in a differentially rotating star can
play a pivotal role in governing mode excitation and damping.

\section{Wave Transport of Angular Momentum} \label{sec:transport}

Let's now flip the record to the `B' side, and discuss what impact
stellar pulsations might have on their host star. Just as waves
transport energy, they also transport momentum. A series of papers by
\cite[Ando (1981, 1982, 1983)]{And1981,And1982,And1983} first
considered how the angular momentum transported by non-axisymmetric
waves alters the internal rotation profile of a star. To model this
process we can perform a Reynolds decomposition of the azimuthal
momentum equation to find the angle-averaged volumetric torque as
\begin{equation} \nonumber
\frac{\partial}{\partial t} \left\langle \varpi \overline{\rho} \overline{v}_{\phi} \right\rangle =
-\frac{1}{4\pi r^{2}} \frac{\partial}{\partial r} L_{J} - 
\frac{\partial}{\partial t} \left\langle \varpi \overline{\rho' v'_{\phi}} \right\rangle -
\left\langle \overline{\rho' \frac{\partial \Phi'}{\partial \phi}} \right\rangle.
\end{equation}
Here $\varpi \equiv \sin\theta$, while the overline denotes the
average over azimuth and the angled brackets the average over
co-latitude. The bracketed term on the left-hand side represents the
angular momentum in a spherical shell of unit thickness. The first
term on the right-hand side is the torque arising from the gradient of
the wave angular momentum luminosity $L_{J}$; the second term is the
rate-of-change of the angular momentum stored in wave motions; and the
third term is the gravitational torque.

Focusing on the first term, the angular momentum luminosity (that is,
the net amount of angular momentum flowing through a spherical surface
in unit time) is given by
\begin{equation} \label{eqn:L_J}
L_{J} = 4 \pi r^{2} \left \langle \varpi \left(\overline{\rho} \overline{v'_{r} v'_{\phi}} +
\overline{v}_{\phi} \overline{\rho' v'_{r}} \right) \right\rangle,
\end{equation}
to second order in the pulsation amplitude \cite[(e.g., Lee \& Saio
  1993)]{LeeSai1993}. The first term in the parentheses is the
Reynolds stress generated by the radial and azimuthal fluid
motions. It vanishes in the case of pure standing waves, because
$v'_{r}$ and $v'_{\phi}$ are exactly 90 degrees out of phase. However,
departures from this strict phase relation arise when waves acquire a
propagative component --- either due to non-adiabatic effects, or from
wave leakage at the outer boundary. In the non-adiabatic case, the
Reynolds stress term for prograde modes transports angular momentum
from excitation regions to dissipation regions (vice-versa for
retrograde modes; see \cite[Ando 1986]{And1986}).

The second term in the parentheses of eqn.~(\ref{eqn:L_J}) is the eddy
mass flux. Again, this term vanishes for pure standing waves, but
becomes non-zero when waves acquire a propagative
component. Shibahashi (these proceedings) proposes the intriguing
hypothesis that the eddy mass flux of g modes, leaking through the
surface layers of Be stars, can transport the angular momentum
necessary to build a quasi-Keplerian disk. The transport is
particularly effective in the outer layers, because the Eulerian
pressure perturbation $\rho'$ is large due to the steep density
gradient there.

Stellar evolution calculations which include wave transport of angular
momentum have so far focused primarily on stochastically excited modes
(see, e.g., Talon 2008 and references therein). However, simple
estimates of transport by overstable global modes suggest that they
can have a significant impact on rotation profiles, over timescales
which are evolutionarily short (Townsend 2008). Interest in this topic
is steadily growing; just this year a number of new papers have
appeared exploring topics such as wave transport in massive stars
\cite[(Rogers et al. 2013)]{Rog2013} and pre-main sequence stars
\cite[(Charbonnel et al. 2013)]{Cha2013}, and the interaction between
wave transport and critical layers \cite[(Alvan et
  al. 2013)]{Alv2013}.

\section{Summary}

To summarize this review, I'd like to highlight an encouraging
trend. Much of the recent progress in understanding the
pulsation-rotation interaction has been driven by numerical
simulations. However there have been multiple parallel efforts to
develop complementary theoretical narratives for the
interaction. These have allowed us to retain a firm grasp on what's
really going on in the simulations, and also reassured us that the
rapidly rotating limit might not be as difficult to understand as we
once thought. Let's ensure this trend does not disappear in the
future, by always remembering the wonderful adage by \cite[Hamming
  (1987)]{Ham1987}: `\emph{the purpose of computing is insight, not
  numbers}'.

\acknowledgements
\noindent I acknowledge support from NSF awards AST-0908688 and
AST-0904607 and NASA award NNX12AC72G. I'd also like to thank the IAU,
and the SOC/LOC, for helping support my travel to this meeting.

\end{document}